\documentclass[12pt,oneside,final,a4paper]{article}
\usepackage{amsmath,amssymb,amsthm,amsfonts}
\usepackage{amsmath,amssymb,makeidx,amsfonts,latexsym,enumerate}

\title{Linear representations of probabilistic transformations induced
by context transitions}

\author{Andrei Khrennikov\\
International Center for Mathematical\\
Modeling in Physics and Cognitive Sciences,\\
MSI, University of V\"axj\"o, S-35195, Sweden\\
Email:Andrei.Khrennikov@msi.vxu.se}

\begin{document}
\maketitle

\begin{abstract}By using straightforward frequency arguments we classify 
transformations of probabilities which
can be generated by transition from one preparation procedure (context) to another. 
There are three classes of transformations
corresponding to  statistical deviations of different magnitudes:
(a) trigonometric; 
(b) hyperbolic; (c) hyper-trigonometric.
It is shown that not
only quantum preparation procedures 
can have trigonometric probabilistic behaviour.
We propose generalizations of ${\bf C}$-linear space 
probabilistic calculus to describe non quantum
(trigonometric and hyperbolic) probabilistic transformations.
We also  analyse superposition principle in this framework.
\end{abstract}

\section{Introduction}

We analyse a well-known expression for probability of an event in terms of 
the conditional probabilities based on another event.
This expression often goes by the name Bayes' formula and is known not to apply in quantum mechanics when 
probabilities for incompatible, or noncommuting, observables are being evaluated, see e.g. [1]-[3].

In the classical case we have Bayes' formula:
\begin{equation}
\label{B}
p(A=a_i)=p(C=c_1) p(A=a_i/C=c_1)+p(C=c_2)p(A=a_i/C=c_2),
\end{equation}
where $A=a_1, a_2$ and $C=c_1, c_2$ 
are two dichotomic random variables. In the quantum case we have the formula:
$$
p(A=a_i)= p(C=c_1) p(A=a_i/C=c_1) + p(C=c_2) 
p(A=a_i/C=c_2)
$$
\begin{equation}
\label{Q}
\pm 2 \sqrt{p(C=c_1) p(A=a_i/C=c_1) p(C=c_2)p(A=a_i/C=c_2)} \cos \theta\; ,
\end{equation}
where $\theta$ is some phase.

The appearance of the interference term in quantum modification of Bayes' formula
has led to the use of the term "quantum probability" in contradiction to what 
could be called "regular" or "classical" probability, see e.g. [1]-[21] for extended
discussions on this problem. But there is only one type of physical 
probability and it is one that is subject to measurement via counting and the generation 
of relative frequencies. It is the relative frequency probability (of von Mises) that is directly
connected with data from experiment. We provide frequency probabilistic analysis making a contribution
to the understanding of probability and Bayes' formula within
the context of quantum mechanics. 

Our analysis begins with the relative frequency definition of the relevant probabilities. The probability
for the eigenvalue of one observable is then expressed in terms of the conditional probabilities 
involving the eigenvalues of a second (in general incompatible) observable. In this way
the "nonclassical" term in quantum Bayes' formula is shown as a perturbation due to the difference 
in preparation procedures of the different states. The perturbing term is then expressed in terms of a coefficient
$\lambda$
whose absolute value can be less or equal to one , or it can be greater than one
for each of the eigenvalues of the observable. This range of values for the coefficient
$\lambda$ then introduces three distinct types of perturbation which are called:
{\it Trigonometric, Hyperbolic, and Hyper-trigonometric.} Each case is then  examined separately.
Classical and quantum cases are then special cases of more general results.
It is then shown that in the quantum case it is possible to reproduce a Hilbert space
in which the probabilities are found in the usual way, but there is a case in which this
is not possible, though the space is linear it is not a Hilbert space. In general it is 
not a complex linear space. In the case of Hyperbolic probabilistic behaviour we have 
to use linear representation of probabilities over so called hyperbolic numbers.

In fact, our approach to experimental probabilities is nothing than the well-known
contextualists approach. In quantum theory such an approach was strongly supported by N. Bohr
[11], H. Heisenberg [3], see also [5]-[21]. Already Heisenberg [3] pointed directly out
that quantum interference of probabilistic alternatives, (\ref{Q}), is a consequence of
transition
from one context (complex of physical conditions) to another. N. Bohr always pointed out that
we have to take into account experimental arrangement to determine "quantum probabilities."
In this paper we did the important step to justify the contextualists approach: we derived
"quantum probabilistic rule" in purely classical frequency probabilistic framework, see also [22], [23].
In fact,
we obtained much more than it was planned. We found that possible modifications of Bayes'
formula (induced by context transitions) are not reduced to "quantum probabilistic rule",
see the above discussion.

As it was already mentioned, it is impossible 
to realize  probabilistic transformations induced by all possible context-transitions in a Hilbert space.
We develop non-Hilbert linear space probabilistic formalism.
The main distinguishing feature of this formalism is {\it the violation
of the superposition principle.} There is no more superposition transitivity: combination of two superpositions
need not be again a superposition. The principle of superposition is the cornerstone of quantum formalism.
There is still the large diversity of opinions on this principle. It may be that our models in that the 
principle of superposition is violated may be useful for analysis of this principle.\footnote{In our models 
situation is not like in models with superselection rules. `Probability superselection' could not be 
represented by choosing a linear subspace in the  Hilbert space of quantum states. In some sense
our selections can be considered as nonlinear `superselections'.} We note that there is a similarity
with quantum formalism based on the theory of POVM (positive operator valued measures), see, for example,
[17], [18], [24], [6], [16]
in that it is possible to consider nonorthogonal expansions of the unit operator. 

The results of this paper were presented in authors talks at International Conferences 
"Foundations of Probability and Physics," V\"axj\"o-2000,  "Quantum Physics: Reconsideration
of Foundations", V\"axj\"o-2001, and "Exploring Quantum Physics", Venice-2001.

\section{Classification of transformation rules for probability distributions for three preparation procedures}
Let ${\cal E}$ be some preparation procedure (see [5]-[10], [13]) 
that produce physical systems having two properties $A$ 
and $C.$ These properties are described by dychotomic variables $A=a_1, a_2$ and $C=c_1, c_2.$ 
We fix the number of preparation acts, $N\equiv N_{{\cal E}}.$ So ${\cal E}$ always
produces ensembles $S=S_{{\cal E}}$ having $N=|S|$ elements. 

Let ${\cal E}_{1}$ and ${\cal E}_{2}$ be two other preparation procedures. It is assumed that
each of these preparation procedures can be applied to elements of $S.$ By application of 
${\cal E}_{i}$ to $S$ we produce a new statistical ensemble $S_i, i=1,2.$ 
\footnote{In general we need two different ensembles $S_{{\cal E}}$ to produce two ensembles, $S_1$ and $S_2.$} 
The main feature of the ensemble $S_i$ is that $C=c_i$ 
for its elements $(i=1,2).$ For example, 
${\cal E}_i$ can be considered as filters with respect 
to the property $C:\; {\cal E}_i$ select elements of $S$ such that $C=c_i(i=1,2).$ 
Such a filtration justifies the assumption that the number of elements 
in $S_i$ could be chosen equal to the number of elements, $N_i, $ 
in $S$ having the property $C=c_i(i=1,2).$ So everywhere below:

$|S_i|=N_i, i=1,2.$

The crucial point of our considerations is that {\bf{in general we could not `select', 
for example, elements with the property $C=c_1$ without  disturbing the property $A.$}}
In general the sub-ensemble
\[S_{ij}=\{s\in S_i:A=a_j\}\] 
of the ensemble $S_i$ does not coincide with the sub-ensemble 
\[S_{ij}^{(0)}=\{s\in S:C=c_i, A=a_j\}\] 
of the original ensemble $S.$ 
We set 
\[n_{ij}=|S_{ij}^{(0)}|\; \; {\rm{and}}\; \; m_{ij}=|S_{ij}|\]
(the numbers of elements in the sub-ensembles) and 
\[N_i=|\{s\in S: C=c_i\}|, n_j=|\{s\in S:A=a_j\}|\] 
(the numbers of elements in $S$ having, respectively, properties
$C=c_i, i=1,2$ and $A=a_j , j=1,2).$  We note that everywhere below the first number,
$i$, in the index pair $ij$ is related to the property $C$ and the second one, $j$, to the property $A.$ 
We shall use the frequency approach to probability (see, [25 ] and [19]): 
the probability is defined as the limit of relative frequencies 
when the number of trials $N \rightarrow \infty.$

{\bf Remark.} (Foundations of Probability and Physics)
{\small As we have already discussed [19], the conventional
probability theory based on Kolmogorov axiomatics [26] is not the best tool 
to work with `quantum probabilities'. The formal use of abstract, absolute, probability
measure is the source of many misunderstandings. In particular, Kolmogorov model
is not the best one for operating with transitions from one context to another. In fact,
all probabilities are conditional probabilities; there is no absolute probability
(see L. Ballentine [27] for the extended discussion). We prefer  to work with frequency probabilities.
Here contexts are described by collectives (random sequences) that are used to find relative frequencies.
However, in this paper we will not pay much attention to the mathematical details
of the frequency framework, see [19] for the details. In fact, everybody who is familiar 
with R. von Mises frequency probability theory could recognize that in this paper 
we work with von Mises collectives. These collectives are produced by different preparation
procedures (complexes of physical conditions). It may be even better to use the term "collective"
instead of the term "ensemble" that is used in this paper. However, we are little bit afraid 
to do this, because there is a rather strong prejudice against von Mises' approach (especially
from the mathematical side).}

We consider relative frequencies: \[q_j^{(N)}\equiv p_j^a(N)=\frac{n_j}{N}, p_i^{(N)}\equiv p_i^c(N)=\frac{N_i}{N} \] (for the properties $A$ and $C$ in the ensembles prepared by $\cal E$);
\[p_{ij}^{a/c}(N_i)\equiv p_{ij}(N_i)=\frac{m_{ij}}{N_i} \](for the property $A=a_j$ in the ensemble prepared by ${\cal E}_i$) and the corresponding probabilities:
\[q_j\equiv p_j^a=\lim_{N\rightarrow \infty} p_j^a(N); p_i\equiv p_i^c=\lim_{N\rightarrow \infty} p_i^c(N); p_{ij}\equiv p_{ij}^{a/c}=\lim_{N_i\rightarrow \infty}p_{ij}(N).\]
As in general $n_{ij}$ do not equal to $m_{ij}$ (even asymptotically $N\rightarrow \infty$), 
we do not have the conventional formula of total probability. In general 

$q_j=p_S(A=a_j)\neq p_1 p_{11} + p_2 p_{21} =
$

$p_S(C=c_1) p_{S_1}(A=a_1)+ p_{S}(C=c_2)p_{S_2}(A=a_2).$

We want to investigate the various forms probabilities $q_j$ can take, depending on
perturbations induced by context transitions. In the general case we have:
\[
q_j(N)=\frac{n_j}{N}=\frac{n_{1j}}{N}+\frac{n_{2j}}{N}=
\frac{m_{1j}}{N}+\frac{m_{2j}}{N}+\delta_j^{(N)} = \frac{N_1}{N}.\frac{m_{1j}}{N_1}+
\frac{N_2}{N}.\frac{m_{2j}}{N_2}+\delta_j^{(N)}=
\]
\[
p_1(N) p_{1j}(N_1)+p_2(N)p_{2j}(N_2)+\delta_j^{(N)}, 
\] 
where the perturbation term (which appears due to the transition from $S$ to $S_1$ and $S_2$)
has the form:
\[
\delta_j^{(N)}\equiv \delta_j ({\cal E}, {\cal E}_1, {\cal E}_2, N) = \frac{1}{N}[(m_{1j}-n_{1j})+(m_{2j}-n_{2j})].
\]
We remark that there exists the limit

$\delta_j=\lim_{N\rightarrow \infty} \delta_j(N)=q_j-(p_1 p_{1j}+p_2 p_{2j}).$

Thus in general we have 
$q_j=p_1 p_{1j}+p_2 p_{2j}+ \delta_j,$
where 

$\delta_j=\lim_{N\rightarrow \infty}\frac{1}{N}[(m_{1j}-n_{1j})+(m_{2j}-n_{2j})]$. 

It is useful to make normalization by setting 

$\delta_j=2\sqrt{p_1p_{1j}p_2p_{2j}} \lambda_j, j=1,2.$ 

The trivial (but important) remark is that there are three possibilities: 

\medskip

(T)$|\lambda_j|\leq 1; $

(H)$|\lambda_j|>1.$

(HT)$|\lambda_1|\leq 1$ and $|\lambda_2|>1$ or $|\lambda_1|>1$ and $|\lambda_2|\leq 1.$

\medskip

In the case (T) we can always represent the coefficient as  $\lambda_j=\cos \theta_j, j=1,2; $ in the case (H) -- 
as $\lambda_j=\pm \cosh \theta_j, j=1,2;$ in the case (HT) -- as $\lambda_1=\cos \theta_1$ and $\lambda_2=
\pm \cosh \theta_2$ or vice versa.
Probabilistic behaviours of the types (T), (H) and (HT) will be called {\it trigonometric, hyperbolic and 
hyper-trigonometric} behaviours, respectively. 

We have studied the general case. 
There are three preparation procedures ${\cal E}, {\cal E}_1$ and ${\cal E}_2$ such
that ${\cal E}_1$ and ${\cal E}_2$ are selections with respect to values $C=c_1$ and $C=c_2.$ 
The general probabilistic transformation induced by transitions ${\cal E} \to {\cal E}_j, j=1,2,$
has the form:
\begin{equation}
\label{G}
p_j^a=p_1^c p_{1j}^{a/c}+p_2^c 
p_{2j}^{a/c}\pm 2\sqrt{p_1^c p_{1j}^{a/c}p_2^c p_{2j}^{a/c}}\lambda_j \;, 
\end{equation}
where
$\lambda_j = \cos \theta_j$ or $\lambda_j= \cosh \theta_j,$ 
or $\lambda_1 =\cos \theta_1$ and $\lambda_2 =\pm \cosh\theta_2$ or vice versa.
Here the coefficient $\lambda_j$ gives the normalized statistical 
measure of the perturbations of $A$ due to the transition ${\cal E}\rightarrow ({\cal E}_1, {\cal E}_2):$
\begin{equation}
\label{ST}
\lambda_j=\lim_{N\rightarrow \infty}\lambda_j^{(N)},\; \mbox{where}\;
\lambda_j^{(N)}=\frac{1}{2\sqrt{m_{1j}m_{2j}}}[n_{1j}-m_{1j})+(n_{2j}-m_{2j})].
\end{equation}

If these perturbations are relatively small, namely $|\lambda_j|\leq 1, j=1,2,$ 
then we observe T-behaviour; in particular, classical and quantum behaviours. 
If these perturbations are relatively large, namely $|\lambda_j| > 1, j=1,2,$ 
then we observe H-behaviour. In fact, we can continuously transfer 
T-behaviour into H-behaviour, since $\lambda_j, |\lambda_j|=1,$ has both T- and H-representations: 
$\lambda_j=\pm \cos 0=\pm \cosh 0.$ If one of these perturbations, for instance $\lambda_1, $ is relatively small, namely $|\lambda_1|\leq 1,$ and another, $\lambda_2,$ is relatively large, namely $|\lambda_2|>1,$ then we observe HT-behaviour.

Finally,  we show that coefficients $\lambda_1$ and $\lambda_2$ 
are connected by a `condition of orthogonality' (in the quantum formalism this is the real 
condition of orthogonality in the complex Hilbert space).
We note that the matrix of probabilities $P =(p_{ij})$ is always a stochastic matrix:
\begin{equation}
\label{S}
p_{11}+p_{12}=1\; \mbox{and}\; \;  p_{21}+p_{22}=1 
\end{equation}
(because $p_{i1}+p_{i2}=p_{S_i}(A={a_1})+p_{S_i}(A=a_2)=1$). 
Thus we have 
$$
1= p_1^a + p_2^a = 
p_1^c p_{11}^{a/c} + p_2^c p_{21}^{a/c} + 
p_1^c p_{12}^{a/c} + p_2^c p_{22}^{a/c}+
$$
$$
2 \sqrt{p_1^c p_{11}^{a/c} p_2^c p_{22}^{a/c}} 
\lambda_1 + 2 \sqrt{p_1^c p_{12}^{a/c} p_2^c p_{22}^{a/c}}\lambda_2 \;.
$$
To simplify considerations, we assume everywhere that all probabilities are strictly positive.
This implies:
\begin{equation}
\label{0}
\sqrt{p_{11}^{a/c} p_{21}^{a/c}} \lambda_1 + \sqrt{p_{12}^{a/c} p_{22}^{a/c} } \lambda_2=0\;.
\end{equation}
 We set
$$K= \sqrt{\frac{p_{12}^{a/c} p_{22}^{a/c}}{ p_{11}^{a/c} p_{21}^{a/c}}}.$$
We get
\[
\lambda_1=-K\lambda_2.
\]
We observe that probabilities $p_j^c$ are not involved in the condition of orthogonality (\ref{0}). In particular, in the T-case we always have:
\begin{equation}
\label{OT}
\cos\theta_1=-K\cos \theta_2
\end{equation}
in the H-case we have:
\begin{equation}
\label{OH}
\cosh \theta_1= K \cosh \theta_2
\end{equation}
(here $\lambda_1= \pm \cosh \theta_1$ and $\lambda_2= \mp \cosh \theta_2).$

In the HT-case we have:
\begin{equation}
\label{OHT}
\cos\theta_1=\pm K \cosh \theta_2\; \mbox{or}\; \cosh \theta_1 =  \pm K \cos\theta_2\;.
\end{equation}

Finally, we remark that all above considerations can be easily generalized to non
dichotomic variables: $A=a_1,...,a_M$ and $C=c_1,...,c_M.$

In this case the probability ${\bf{p}}_i^a$ can always be represented as 
\begin{equation}
\label{LA}
{\bf{p}}_i^a=\sum_{j=1}^M {\bf{p}}_j^c {\bf{p}}_{ji}^{a/c} + 
2\sum_{k<l}\sqrt{{\bf{p}}_k^c {\bf{p}}_l^b {\bf{p}}_{ki}^{a/c} {\bf{p}}_{li}^{a/c}} \lambda_{kl}^{(i)},
\end{equation}
 where the coefficients 
 $\lambda_{kl}^{(i)}=\frac{\delta_{kl}^{(i)}}{2\sqrt{{\bf{p}}_k^c {\bf{p}}_l^c {\bf{p}}_{ki}^{a/c} {\bf{p}}_{li}^{a/c}}}$
 
 and
$\delta_{kl}^{(i)}=\frac{1}{M-1}[{\bf{p}}_k^c ({\bf{p}}_i^a - {\bf{p}}_{ki}^{a/c}) + 
{\bf{p}}_l^c ({\bf{p}}_i^a - {\bf{p}}_{li}^{a/c})]$
Coefficients \{$\lambda_{kl}^{(i)}$\} are normalized statistical 
deviations that arise due to the transition from the context determined 
by ${\cal{E}}$ to contexts ${\cal{E}}_j.$ 

To simplify analysis, we shall consider only dichotomic variables in the following sections.

\section{Trigonometric probabilistic behaviour: classical, quantum and non classical/quantum physics}

In this section we consider probabilistic transformations for preparation procedures that produce
relatively small statistical deviations:
$$
\vert \lambda_j \vert\leq 1, j=1,2.
$$

{\bf{1. Classical probabilistic behaviour.}}
Suppose that we can construct statistically `perfect' preparation procedures ${\cal E}_1, 
{\cal E}_2:$ selections of elements of the ensemble $S$ with respect to values $C=c_1$ and $C=c_2$ 
produce statistically negligible changes of $A.$ We set

$\Delta_{ij}(N)=n_{ij}-m_{ij}.$

Here $n_{ij}$ is the number of elements of $S$ having $C=c_i$ and $A=a_j$ and $m_{ij}$ is the number of elements of $S_i$ having $A=a_j.$ The classical probabilistic behaviour is characterized by the condition: 
\[
\lim_{N\rightarrow \infty}\frac{\Delta_{ij}(N)}{N}=0, \; \mbox{for all}\; i,j.
\] 
Here both $\lambda_j=0$ and we have conventional rule (\ref{B}). 

{\bf{2. Quantum probabilistic behaviour.}}
Let us consider preparations which induce symmetric statistical deviations:
\begin{equation}
\label{SS}
|\lambda_1|=|\lambda_2| \;.
\end{equation}

Thus the coefficient $K$ is equal to 1. So
$p_{12}p_{22}=p_{11}p_{21}.$
In the two dimensional case this condition is equivalent to the well known 
condition of {\bf{double stochasticity}}:
\begin{equation}
\label{DS}
p_{11} +p_{21}=1, p_{12}+p_{22}=1\;.
\end{equation}
Thus $p_{S_1}(A=a_1)+p_{S_2}(A=a_1)=1$ and $p_{S_1}(A=a_2)+p_{S_2}(A=a_2)=1.$ These are `conservation laws' 
for the $A$ in the process of splitting of the ensemble $S$ into ensembles $S_1$ and $S_2$.
We also remark that (\ref{OT}) implies that $\cos \theta_1=-\cos \theta_2.$ So $\theta_2=\theta_1+\pi \; (\mod 2\pi).$ 
Thus we have the probabilistic transformations:
\begin{equation}
\label{QQ}
q_1(\equiv p_1^a)=p_1 p_{11}+p_2 p_{21}+2\sqrt{p_1 p_{11}p_2 p_{21}}\cos \theta \;,
\end{equation}
\begin{equation}
\label{QQQ}
q_2(\equiv p_2^a)=p_1 p_{12}+p_2 p_{22}-2\sqrt{p_1 p_{12}p_2 p_{22}}\cos \theta\; .
\end{equation}
This is the well known quantum probabilistic transformation.
We now find complex representations of these probabilities that would linearize transformations
(\ref{QQ}), (\ref{QQQ}).
We use the well known formula:
\begin{equation}
\label{R}
A+B \pm2\sqrt{AB}\cos\theta = |\sqrt{A}\pm \sqrt{B}e^{i\theta}|^2. 
\end{equation}
Thus 

$q_1=|\sqrt{p_1}\sqrt{p_{11}} + \sqrt{p_2}\sqrt{p_{21}}e^{i\theta_1}|^2;
q_2=|\sqrt{p_1}\sqrt{p_{21}} + \sqrt{p_2}\sqrt{p_{22}}e^{i\theta_2}|^2.
$

(in quantum case $\theta_1 = \theta_2 +\pi).$
These formulas can be also derived by ${\bf C}$-linear space computations. 
We represent the preparation procedure ${\cal E}$ by a vector $\varphi$ in the two dimensional complex Hilbert space:

$\varphi=\sqrt{p_1}\varphi_1+\sqrt{p_2}e^{i\theta}\varphi_2\;.$

where $\{\varphi_1, \varphi_2\}$ is an orthonormal basis corresponding to the 
physical observable $C$ (the condition $p_1+p_2=1$ implies that $||\varphi||^2=1).$ 
Let $\psi_1, \psi_2$ be an orthonormal basis corresponding to the 
physical observable $A.$ We have:

$\varphi_1=\sqrt{p_{11}}\psi_1 + e^{i\gamma_1}\sqrt{p_{12}}\psi_2,\;
\varphi_2=\sqrt{p_{21}}\psi_1 +e^{i\gamma_2}\sqrt{p_{22}}\psi_2.$

We remark that orthogonality of $\varphi_1$ and $\varphi_2$ is,
in fact, equivalent to the condition of double stochasticity for $P=(p_{ij})$ and the relation 
$\gamma_2=\gamma_1+\pi \; (\mod 2\pi).$ 
 By expanding $\varphi$ with respect to the basis $\{\psi_1, \psi_1\}$ we get 
 
 $
 \varphi=d_1\psi_1+d_2\psi_2, 
 $
 
 where 
 \begin{equation}
 \label{LT}
 d_1=\sqrt{p_1}\sqrt{p_{11}}+e^{i\theta}\sqrt{p_2 p_{21}}, d_2=e^{i\gamma_1}\sqrt{p_1}\sqrt{p_{12}}+
 e^{i(\gamma_2+\theta)}\sqrt{p_2 p_{22}}\;.
 \end{equation}
 
By using the relation $\gamma_2=\gamma_1+\pi$ we reproduce quantum probabilistic rule
(\ref{QQ}), (\ref{QQQ}). 

We note that our considerations demonstrated that the main distinguishing feature
of quantum formalism is not the presence of 
$\cos\theta$-factor in the `quantum transformation of probabilities,' but the double stochasticity of 
the matrix $P=(p_{ij}^{a/c})$ of transition probabilities and the relation 
\begin{equation}
\label{f}
\gamma_2=\gamma_1+\pi
\end{equation}
between phases in the expansions of $\varphi_1$ and $\varphi_2$ with respect to the basis $\{\psi_1, \psi_2\}$

The `double stochasticity conservation laws', (\ref{DS}), and the `phase conservation law', (\ref{f})  imply the unitarity of the transformation $U$ connecting $\{\varphi_1, \varphi_2\}$ and $\{\psi_1, \psi_2\}.$ In fact, this is the root of the {\bf{superposition principle}} (see the next subsection for the details).

Finally, we remark that there is the crucial difference between 
classical physical behaviour $(\lambda_1=\lambda_2=0$) and 
quantum decoherence $(\lambda_1=\lambda_2=0.)$ In the first case coefficients 
$\lambda_j=0,$ because statistical deviations are negligibly small. In the second case 
coefficients $\lambda_j=0,$ because statistical deviations  compensate each other $(j=1,2):$
\[
\frac{\Delta_{1j}}{N}\approx - \frac{\Delta_{2j}}{N}, N\rightarrow \infty.
\]

{\bf{3. Non classical/quantum  trigonometric probability behaviour.}}
Here the matrix $P =(p_{ij})$ of transition probabilities need not be
not double stochastic. We can find the probability distribution 
$q_j=p_j^a=p_{S}(A=a_j), j=1,2,$ by using the following transformation of probabilities: 
\begin{equation}
\label{eq}
q_j=p_1p_{1j}+p_2p_{2j}+ 2\sqrt{p_1p_{1j}p_2p_{2j}}\cos\theta_j,
\end{equation}
where $\cos\theta_1=-K\cos\theta_2, K=\sqrt{\frac{p_{12}p_{22}}{p_{11}p_{21}}}.$ 
In general such a probabilistic transformation (`interference' between preparation procedures 
${\cal E}_1$ and ${\cal E}_2)$ could not be described by standard quantum formalism. 

{\bf{Example 3.1.}} {\small Let $p_1=p_2=\frac{1}{2}$ (symmetric distribution of $C$ in $S$; for example,
the two slit experiment with symmetric location of slits with respect to the source of particles) 
and 
let $p_{11}=p_{12}=\frac{1}{2}$ (symmetric distribution of $A$ in $S_1$) and 
$p_{21}=\frac{1}{3}, p_{22}=\frac{2}{3}$ (asymmetric distribution of $A$ in $S_2$).
 Thus the matrix $P$ is not double stochastic.

The law of conservation of the $A$ is violated in the process of 
the transition $S\rightarrow (S_1, S_2).$ The measure 
of this violation is given by the coefficient $K.$ Here $K=\sqrt{2}.$ 
Phases  $\theta_1$ and $\theta_2$ must be chosen in such a way that
$\cos\theta_1=-\sqrt{2}\cos\theta_2.$ For example, we can consider preparations such that 
$\theta_1=\frac{3\pi}{4}$ and $\theta_2=\frac{\pi}{3}.$ In this case we have 

$p_1^a=\frac{5}{12}+\frac{\cos \frac{3\pi}{4}}{\sqrt{6}}; p_2^a=\frac{7}{12}+\frac{\cos \frac{\pi}{3}}{\sqrt{3}}. $

This probabilistic transformation could not be obtained in standard `quantum linear calculus'. 
We shall see that it could be obtained by non-unitary generalization of `quantum linear calculus'.} 

\section{Hyperbolic probabilistic behaviour}
In this section we consider examples of H-behaviour and HT-behaviour.  We remark that H-behaviour can be
exhibited by preparations having double stochastic transition matrixes. 

{\bf{Example 4.1.}} {\small Let $p_1=\alpha$ and $p_2=1-\alpha \; 
(0<\alpha<1) $ and let $p_{ij}=1/2, i, j=1,2.$ Here $K=1$ (the transition matrix is double stochastic)
and, hence, $\cosh \theta_2 = \cosh \theta_1.$ We have 
\[q_1=\frac{1}{2}+ \sqrt{\alpha(1-\alpha)} \cosh \theta, \]
\[q_2=\frac{1}{2}- \sqrt{\alpha(1-\alpha)} \cosh \theta, \]
In the opposite to the T-case the phase 
$\theta$ cannot take arbitrary values.  
There is a relation between $\theta$ and $\alpha$ that provides 
that $q_1, q_2$ 
have the meaning of probabilities. We set \[e(\alpha)=\frac{1}{2\sqrt{\alpha(1-\alpha)}}.\]

We remark that $e(\alpha) \geq 1$ for all $0<\alpha<1.$ 
The hyperbolic phase  $\theta$ can be chosen as $\theta \in [ 0, \theta_{\max}],$ 
where $\theta_{\max} = \rm{arccosh}\; e(\alpha).$ For example, 
let $\alpha=\frac{1}{4}(1-\alpha=3/4).$ Thus $e(x)=\frac{2}{\sqrt{3}}.$ Here 
we could observe hyperbolic interference for angles 
$0\leq\theta \leq  \rm{arccosh}\; \frac{2}{\sqrt{3}}.$ We remark that if 
$p_1=p_2=\frac{1}{2},$ then $e(\alpha)=1$ and the hyperbolic interference 
coincides with the ordinary interference $\cos 0 = \cosh 0 = 1.$ 
In general the symmetric distribution $p_1 = p_2 = 1/2$
can produce nontrivial hyperbolic interference. We have for general double stochastic
matrix $p:$

$q_j=\frac{1}{2}(p_{1j}+p_{2j})+\sqrt{p_{1j} p_{2j}}\lambda_j
=\frac{1}{2}+\sqrt{p_{1j} p_{2j}}\lambda_j=\frac{1}{2}+\sqrt{\alpha(1-\alpha)}\lambda_j,$ 

where we set 
$\alpha=p_{11}=p_{22}$ and $1-\alpha=p_{12}=p_{21}.$ If $\theta \in [0, \theta_{\max}], 
\theta_{\max}= \rm{arccosh}\; e(\alpha),$ 
then $\lambda_j = \pm \cosh \theta, \theta \neq 0.$ }

We remark that the total symmetry (in $S$ as well as $S_1, S_2),$ 
namely $p_1= p_2= p_{ij}=1/2,$ produces the trivial H-interference
(that coincides with the T-interference). So hyperbolic interference might be observed only for
preparation procedures with asymmetric probability distributions for contexts.

{\bf Remark.} (Negative probabilities) {\small If we do not pay attention to the 
 range of the H-phase parameter $\theta$ we could get {\it negative probabilities}
and probabilities $>1.$ It must be noted that such `probabilities' appear with the
intriguing regularity in various extensions of quantum formalism (Wigner [28], Dirac [29],
Feynman [30], see also [19]] for the details). It may be that `quantum negative probabilities'
have the same origin as `negative H-probabilities,' namely the use of nonphysical values of some 
parameters, see [19] for the details.}

Of course, our considerations induce the following natural question: 
`Is it possible to construct a linear space representation for
the H-probabilistic transformations?' We shall study this question in section 6.

Finally, we consider an example of mixed HT-behaviour. 

{\bf{Example 4.2.}} 
{\small Let $p_1=p_2=\frac{1}{2}$ and let 
$p_{11}=\frac{4}{5}, p_{12}=\frac{1}{5}, p_{21}=\frac{4}{5}, p_{22}=\frac{1}{5}.$ We have $K=\frac{1}{4};$ so $\lambda_2=-4\lambda_1.$

We have $q_1=\frac{4}{5}(1+\lambda_1), q_2=\frac{1}{5}(1-4\lambda_1).$ If $-1\leq \lambda_1\leq\frac{1}{4},$ 
then $q_1$ and $q_2$ have the meaning of probabilities. 
For example, let $\lambda_1=\frac{-1}{2}$ and $\lambda_2=2.$ Then $q_1=\frac{2}{5}, q_2=\frac{3}{5}.$ 
Thus }

$q_1=\frac{4}{5}+\frac{4}{5}\cos \frac{2}{3}\pi, q_2=\frac{1}{5}+\frac{1}{5}\cosh (\ln(2+\sqrt{3})).$

We remark that mixed HT-behaviour can not be produced on the basis of  a double stochastic matrix $P=(p_{ij}).$

Finally, we note that the H-phase has a symmetry, $\theta\to - \theta,$
that is an analogue of the symmetry $\theta\to \theta + 2\pi$ for the T-phase.
If $\lambda = \cosh \theta,$ then $\theta$ can be chosen as 

$\theta=\ln( \lambda + \sqrt{\lambda^2-1})$ or $\theta= \ln( \lambda -\sqrt{\lambda^2-1})\;.$

\section{Complex linear space representation of the general trigonometric probabilistic rule}

We shall study the possibility to represent general probabilistic transformation (\ref{eq}) 
as a linear transformation in a complex linear space. As in general the transition probability 
matrix $P = (p_{ij})$ is not double stochastic, we could not expect that it would be possible 
to work with orthonormal bases in a complex Hilbert space. 
It seems that the inner product structure is not useful in the general case. 

Let E be a two dimensional linear space over the field of complex numbers $\bf{C}.$ 
The choice of $\bf{C}$ as the basic number field has the trivial explanation.
Formula (\ref{R}) gives the possibility to represent the T-probabilistic transformation
in form (\ref{LT}) which is reduced to the transition from one basis to another.
It is impossible to linearize quantum probabilistic transformation by using
real numbers, but it is possible to do this by using complex numbers.
These arguments were already evident in our analysis of quantum theory. 
We now observe that they can be used in more general situation.

Vectors of $E$ are said to be quantum states. At the moment there is no Hilbert
structure on $E.$ There is no anything similar to the standard normalization condition
for quantum states. We represent the ensemble $S$ 
(the preparation procedure ${\cal E}$) by a vector 
$\varphi$ in $E;$ the ensembles $S_1$ and $S_2$ (the preparation procedures 
${\cal E}_1$ and ${\cal E}_2$) - by vectors $\varphi_1$ and $\varphi_2.$

It is supposed that the preparation procedures ${\cal E}_1$ and ${\cal E}_2$
determine some dichotomic physical variable, $C=c_1,c_2.$ In the linear space calculus
this assumption has the following counterpart: vectors 
$\{\varphi_1, \varphi_2\}$ are linearly independent in $E.$

Splitting of $S$ into $S_1$ and $S_2$ (due to the preparation procedures 
${\cal E}_1$ and ${\cal E}_2$) is represented as expending of the vector
$\varphi$ with respect to a basis $\{\varphi_1, \varphi_2\}$ in $E.$
We can always expend the vector $\phi$ with respect to the basis:

$\varphi=\alpha_1\varphi_1+\alpha_2\varphi_2,$

where $\alpha_1$ and $\alpha_2\in {\bf{C}}.$ As in the ordinary quantum formalism
the probabilities $p_i^c= P_S (C=c_i)$ are represented as $p_i^c=|\alpha_i|^2$
(generalization of Born's postulate). So there is a constraint for vectors
$\varphi$ and $\varphi_1$, $\varphi_2:$
\begin{equation}
\label{fi1}
|\alpha_1|^2+|\alpha_2|^2=1\;.
\end{equation}
In such a case the quantum state $\varphi$ is said to be $C$-{\it decomposable.}

We now consider the measurement of $A$ for ensembles 
$S_i$ (prepared by ${\cal E}_i).$ We consider such a measurement 
that a second measurement of $A$, performed immediately after the first one, 
will yield the same value of the observable. In quantum theory such measurements are 
often called {\it `measurements of the first kind'}. Thus such 
$A$-measurement can be interpreted as a preparation procedure. 

To be more precise, we consider two preparation procedures ${\cal E}_1^a$ and ${\cal E}_2^a$ 
corresponding to selections of physical systems on the basis of values $A=a_1$
and $A=a_2.$  The $C$-preparation procedures ${\cal E}_1$ and ${\cal E}_2$ we now denote by the symbols 
${\cal E}_1^c$ and ${\cal E}_2^c,$ respectively. 
${\cal E}_j^a$ selects physical systems such that $A=a_j, j=1,2.$ We remark 
that in general these selections may change the probability distribution of $C$. 
By applying ${\cal E}_j^a$ to the ensemble $S_i^c\equiv S_i$ (which was produced by 
the application of ${\cal E}_i^c$ to an ensemble $S$ produced by ${\cal E}$) 
we obtain an ensemble $S_{ij}^{ca}, i, j=1,2.$ In the same way we split 
the ensemble $S$ (with the aid of ${\cal E}_1^a$ and ${\cal E}_2^a$) into ensembles 
$S_j^a, j=1,2.$
Ensembles $S_j^a, j=1,2,$ are represented by vectors $\psi_j$ in the $E.$ 
We assume that they also form a basis in $E$ (this is a consequence of the fact
that preparation procedures ${\cal E}_1^a$ and ${\cal E}_2^a$ determine the 
dichotomic physical variable $A).$
Thus splitting $S \rightarrow (S_1^a, S_2^a)$ can be represented by the expansion 

$\varphi= \beta_1 \psi_1+\beta_2 \psi_2,$

where $\beta_j\in {\bf{C}}.$ Here probabilities $p_j^a= P_S(A=a_j)=|\beta_j|^2,$ so 
\begin{equation}
\label{fi2}
|\beta_1|^2+|\beta_2|^2=1.
\end{equation}
Thus $\varphi$ is $A$-decomposable.

In the general case we have to represent ensembles $S_{ij}^{ca}, i, j=1,2,$ by four 
different vectors $\psi_{ij}.$ In general we cannot assume that 
these vectors belong to the same two-dimensional space $E.$ 
The study of this general situation is too complicated. 
We restrict ourself to the special case (which is the most interesting for applications). 
Let $\psi_{11}=\psi_{1}$ and $\psi_{21}=\psi_{1}, \psi_{21}=\psi_{2}$ and $\psi_{22}=\psi_{2}.$
It was assumed that $\psi_1$ and $\psi_2$ are independent vectors. 

We would like to predict the probabilities $p_j^a$ on the basis of 
the transition from the basis $\{\varphi_1, \varphi_2\}$ to the basis $\{\psi_1, \psi_2\}.$ 
Let $U= (\beta_{ij})$ be the transition matrix (the only restriction to $U$ is its invertibility).
Here each vector $\varphi_i$ is $A$-decomposable.\footnote{In general there is no composition (or it would be better
to say decomposition) transitivity. For example, it may be that the state $\varphi$ is
$C$-decomposable and each state $\varphi_i$ is $A$-decomposable, but $\varphi$ is not $A$-decomposable.
We suppose decomposability of all states under the consideration by physical reasons: the possibility
to perform $A$ and $C$ measurements for elements of $S.$ The violation of composition transitivity
corresponds to the following situation: we can perform $C$-measurement  on $S$ and $A$-measurements
on $S_i^c$, but we could not perform $A$-measurement on $S.$} Thus
\begin{equation}
\label{ffi2}
|\beta_{i1}|^2+|\beta_{i2}|^2=1, i=1,2.
\end{equation}
We have
\begin{equation}
\label{TR}
\beta_1=\alpha_1\beta_{11}+\alpha_2\beta_{21}, \beta_2=\alpha_1\beta_{12}+\alpha_2\beta_{22}.
\end{equation}
Coefficients $\alpha_j, \beta_{ij}$ are not independent. 
They satisfy to constraint (\ref{fi2}). Simple computations give us 
\begin{equation}
\label{CON}
\alpha_1\bar{\alpha}_2(\beta_{11}\bar{\beta}_{21}+\beta_{12}\bar{\beta}_{22})+ \bar{\alpha}_1{\alpha}_2(\bar{\beta}_{11}\beta_{21}+\bar{\beta}_{12}\beta_{22})=0.
\end{equation}
One of solutions of this equation is given by
\begin{equation}
\label{CON1}
\beta_{11}\bar{\beta}_{21}+\beta_{12}\bar{\beta}_{22}=0.
\end{equation}
This is the condition of unitarity of the transition matrix $U=(\beta_{ij}).$ 
This solution gives ordinary quantum formalism. In this formalism it is useful
to introduce the inner product: 
\[< z, w > =z_1\bar{w}_1+z_2\bar{w}_2\] 
and rewrite the above equation as the condition of orthogonality of vectors $\varphi_1$ and $\varphi_2:$
$ < \varphi_1, \varphi_2>=0.$
However, equation (\ref{CON}) 
have other solutions which are not related to standard quantum formalism. 
These solutions give the complex linear space representation 
for the trigonometric probabilistic rule in the non classical/quantum case. We set:

$\alpha_i=\sqrt{p_i}e^{i\xi_i}, \beta_{ij}=\sqrt{p_{ij}} e^{i\gamma_{ij}},$

where $p_1+p_2=1, p_{11}+p_{12}=1, p_{21}+p_{22}=1$ and $\xi_1, \gamma_{ij}$ are arbitrary phases. 
Thus the transition from one basis to another has the form:
\begin{equation}
\label{NU}
\varphi_1= \sqrt{p_{11}} e^{\gamma_{11}} \psi_1+ \sqrt{p_{12}} e^{\gamma_{12}} \psi_2\;,\;
\varphi_2= \sqrt{p_{21}} e^{\gamma_{21}} \psi_1+ \sqrt{p_{22}} e^{\gamma_{22}} \psi_2\;.
\end{equation}
In these notations equation (\ref{CON}) has the form:
\begin{equation}
\label{OR1}
\cos(\eta + \gamma_1)\sqrt{p_{11}p_{21}}+\cos(\eta + \gamma_2)\sqrt{p_{12}p_{22}}=0, 
\end{equation}
where $\eta=\xi_1-\xi_2, \gamma_1=\gamma_{11}-\gamma_{21}, \gamma_2=\gamma_{12}-\gamma_{22}.$

We set $\theta_1=\eta+\gamma_1$ and $\theta_2=\eta+\gamma_2.$ Equation (\ref{OR1}) 
coincides with equation (\ref{0}) in the T-case. Thus all possible probabilistic 
T-transformations can be represented in the complex linear space. A rather surprising 
fact is that equation (\ref{OR1}) has a new (nonquantum solution) even for a 
double stochastic matrix of transition probabilities. 

Let $P$ be a double stochastic matrix. Equation (\ref{OR1}) has the form: 

$\cos(\eta+\gamma_1)+\cos(\eta+\gamma_2)=0\;.$

Thus

$\cos \frac{(2\eta+\gamma_1+\gamma_2)}{2}=0$ or $\cos\frac{(\gamma_1-\gamma_2)}{2}=0$

There is the crucial difference between these equations.
The first equation `remembers' the state $\varphi;$  splitting of
$\varphi$ into $\{\varphi_1, \varphi_2\}$
(or $S$ into $S_1$ and $S_2$). This memory is given by the phase shift $\eta.$ 
The second equation does not contain any memory term. In fact, 
this is the standard quantum mechanical equation: $\gamma_1-\gamma_2=\pi \; (\mod 2 \pi).$ 

Thus we get a new (nonquantum) solution even for a double stochastic matrix 
$P=(p_{ij}):$

$2\eta + \gamma_1 + \gamma_2=\pi\; (\mod 2\pi).$

In this case transformation (\ref{NU}) also reproduce quantum probabilistic rule
(\ref{QQ}), (\ref{QQQ}): $q_j=p_1p_{1j}+p_2p_{2j}\pm  2\sqrt{p_1p_{1j}p_2p_{2j}} \cos\theta.$
However, (\ref{NU}) is not unitary: 

$
\beta_{11} \bar{\beta}_{21} + \beta_{12} \bar{\beta}_{22}= 1 - e^{-2i\eta}\not= 0, \eta\not=0.
$

\section{Linear space representation of the hyperbolic probabilistic rule}
We want to find a kind of linear space calculus for 
the H-probabilistic transformation. It seems that it 
would be impossible to do this in a ${\bf{C}}$-linear space. We propose 
to use a hyperbolic algebra {\bf{G}}, see [31]. This is a two dimensional real algebra 
with basis $e_0 = 1$ and $e_1 = j, $ where $j^2=1.$

Elements of {\bf{G}} have the form $z=x + j y, \; x, y \in {\bf{R}}.$ 
We have $z_1 + z_2=(x_1+x_2)+j(y_1+y_2)$ and $z_1 z_2=(x_1x_2+y_1y_2)+j(x_1y_2+x_2y_1).$ 
This algebra is commutative. We introduce an involution in {\bf{G}} by setting 
$\bar{z} = x - j y.$ 
We set 

$|z|^2=z\bar{z}=x^2-y^2.$ 

We remark that  $|z|=\sqrt{x^2-y^2}$ is not well defined for an arbitrary $z\in {{\bf{G}}}.$ 
We set ${{\bf{G}}}_+=
\{z\in{{\bf{G}}}:|z|^2\geq 0\}.$ We remark that ${{\bf{G}}}_+$ 
is the multiplicative semigroup: 
$z_1, z_2 \in {{\bf{G}}}^+ \rightarrow z=z_1 z_2 \in {{\bf{G}}}_+.$ 
It is a consequence of the equality 

$|z_1 z_2|^2=|z_1|^2 |z_2|^2.$

Thus, for $z_1, z_2 \in {{\bf{G}}}_+,$ 
we have $|z_1 z_2|=|z_1||z_2|.$ We introduce 
$$
e^{j\theta}=\cosh\theta+ j \sinh\theta, \; \theta \in {\bf{R}}.
$$ 
We remark that 

$e^{j\theta_1} e^{j\theta_2}=e^{j(\theta_1+\theta_2)}, \overline{e^{j\theta}} 
=e^{-j\theta}, |e^{j\theta}|^2= \cosh^2\theta - \sinh^2\theta=1.$

Hence, $z=\pm e^{j\theta}$ always belongs to ${{\bf{G}}}_+.$ 
We also have 
$$
\cosh\theta=\frac{e^{j\theta}+e^{-j\theta}}{2}, \;\;\sinh\theta=\frac{e^{j\theta}-e^{-j\theta}}{2 j}\;.
$$

We set ${{\bf{G}}}_+^*=
\{z\in{{\bf{G}}}_+:|z|^2>0 \}. $ 
Let  $z\in {{\bf{G}}}_+^*.$  We have 

$z=|z|(\frac{x}{|z|}+j \frac{y}{|z|})= \rm{sign}\; x\; |z|\;(\frac{x {\rm{sign}} x}{|z|} +j\;
\frac{y {\rm{sign}} x}{|z|}).$

As $\frac{x^2}{|z|^2}-\frac{y^2}{|z|^2}=1,$  we can represent $x$ sign $x= \cosh\theta$ 
and $y$ sign $x=\sinh\theta, $ where the phase $\theta$ is unequally defined. 
We can represent each $z\in {{\bf{G}}}_+^*$ as 

$z = \rm{sign}\; x\;  |z|\; e^{j\theta}\;.$ 

By using this representation we can easily prove that ${{\bf{G}}}_+^*$
is the multiplicative group. Here $\frac{1}{z}=\frac{{\rm{sign}} x}{|z|}e^{-j\theta}.$ 
The unit circle in ${{\bf{G}}}$ is defined as $S_1 = \{z\in{{\bf{G}}}:|z|^2=1\}
=\{ z= \pm e^{j \theta}, \theta \in (-\infty, +\infty)\}.$ It is a multiplicative
subgroup of ${\bf G}_+^*.$

Hyperbolic Hilbert space is 
${{\bf{G}}}$-linear space (module) ${\bf{E}}$ 
with a ${{\bf{G}}}$-scalar
product: a map $(\cdot, \cdot): {\bf{E}}\times {\bf{E}} \to {{\bf{G}}}$ that is 

1) linear with respect to the first argument: 

$ (a z+ b w, u) = a (z,u) + b (w, u), a,b \in {{\bf{G}}}, 
z,w, u \in {\bf{E}};$

2) symmetric: $(z,u)= \overline{(u,z)} ;$

3) nondegenerated: $(z,u)=0$ for all $u \in {\bf{E}}$ iff $z=0.$

We note that 1) and 2) imply that 

$(u, a z+ b w) = \bar{a} (u,z) + \bar{b} (u, w).$

{\bf Remark.} If we consider ${\bf{E}}$ as just a ${\bf R}$-linear space, then $(\cdot, \cdot)$ 
is a (rather special) bilinear form which is not positively defined.  
In particular, in the two dimensional case we have the signature: $(+,-,+,-).$

We shall represent the H-probabilistic transformation in 
the two dimensional ${{\bf{G}}}$-linear space (module) ${\bf{E}}.$ 
From the beginning we do not consider any
${\bf{G}}$-Hilbert structure on ${\bf{E}}.$ Such a structure will appear 
automatically in the representation of one particular class of H-probabilistic 
transformations, H-quantum formalism. In the same way as in the previous section
we introduce quantum states $\varphi, \{\varphi_1, \varphi_2\}, \{\psi_1, \psi_2\}$
corresponding to preparation procedures (statistical ensembles). By definition 
a quantum state is a vector belonging to a ${\bf G}$-linear space (no normalization!).

It is supposed
that $\{\varphi_1, \varphi_2\}$ and  $\{\psi_1, \psi_2\}$ are bases in the ${\bf G}$-linear
space $E.$ 

It is supposed that the state $\varphi$ is $C$ and $A$-decomposable and the states 
$\varphi_i$ are $A$-decomposable. Thus:

$\varphi= \alpha_1 \varphi_1 + \alpha_2\varphi_2, |\alpha_1|^2+|\alpha_2|^2=1, |\alpha_j|^2\geq 0,$ 

and 

$\varphi_1=\beta_{11}\psi_1+\beta_{12}\psi_2, \varphi_2= \beta_{21}\psi_1+\beta_{22}\psi_2,$ 

where vectors of coefficients $\beta^{(1)}=(\beta_{11}, \beta_{12})$ and $\beta^{(2)}=(\beta_{21}, \beta_{22})$ 
are such that 

$|\beta_{11}|^2+|\beta_{12}|^2=1|\beta_{21}|^2+|\beta_{22}|^2=1$ and $|\beta_{ij}|^2\geq 0.$

Thus 

$\varphi=\beta_1\psi_1+\beta_2\psi_2,$

where the coefficients $\beta_1, \beta_2$ are given by (\ref{TR}).
There is no formal difference between linear space transformations over ${\bf C}$ and ${\bf G}.$
However, the assumption that the state $\varphi$ is $A$-decomposable implies that
the ${\bf G}$-linear space calculations have a physical meaning iff 
the vector $\beta=(\beta_1, \beta_2)$ is such that
\begin{equation}
\label{E1}
|\beta_1|^2=|\alpha_1\beta_{11}+\alpha_2\beta_{21}|^2\geq 0,| \beta_2|^2=|\alpha_1\beta_{12}+\alpha_2\beta_{22}|^2\geq 0,
\end{equation}
and 
\begin{equation}
\label{NT}
|\beta_1|^2+|\beta_2|^2=1.
\end{equation}
The latter equation coincides with equation (\ref{CON}) 
(with the only difference that all numbers belong to {\bf{G}} instead of {\bf{C}}).

As we have already discussed in the T-case, in general 
there is no composition (in fact, decomposition) transitivity.
In general the $C$-decomposability of $\varphi$ and $A$-decomposability of 
$\varphi_i$ need not imply that $\varphi$ is also $A$-decomposable. Our assumptions
on composition transitivity are based on the physical context of our considerations.

As in the {\bf{T}}-case, (\ref{CON}) has the solution given by 
equation (\ref{CON1}) (the only difference is that now all coefficients belong to
the hyperbolic algebra). This is the condition of 
orthogonality of vectors $\varphi_1$ and $\varphi_2$ with respect to the 
{\bf{G}}-linear product:$ < z, w > =z_1\bar{w}_1+z_2\bar{w}_2.$ 
So the matrix $U =(\beta_{ij})$ is a {\bf{G}}-unitary matrix, namely
\begin{equation}
\label{W}
< \beta^{(i)}, \beta^{(j)} >=\delta_{ij}\;.
\end{equation}

We now study the general case. Here the $U$ need not be unitary matrix.
We consider only vectors with coefficients belonging to ${\bf G}_+^*.$
We set $\alpha_i=\pm\sqrt{p_i} e^{j\xi_i}, \beta_{ij}=\pm \sqrt{p_{ij}} e^{j\gamma_{ij}}, i, j, =1,2.$ 
Condition (\ref{NT}) is equivalent to the condition: 

$\sqrt{p_{12}p_{22}} \cosh \theta_2 + \sigma \sqrt{p_{11}p_{21}} \cosh \theta_2= 0\;,$

where $\sigma=\sqcap_{ij} {\rm{sign}} \beta_{ij}.$ This equation has a solution, namely phases 
$\theta_1$ and $\theta_2,$ iff
\begin{equation}
\label{SI}
\sigma =-1.
\end{equation}
Thus the transition matrix U=$(\beta_{ij})$ must always satisfy (\ref{SI}). 

Let us turn back to the case in that $U$ is a {\bf{G}}-unitary matrix. 
We shall call such a model {\it hyperbolic quantum formalism.} The orthogonality relation implies: 

$0=(\beta^{(1)}, \beta^{(2)})=$

${\rm{sign}} \beta_{11} {\rm{sign}} \beta_{21}\sqrt{p_{11}p_{21}}
e^{j(\gamma_{11}-\gamma_{21})}+ {\rm{sign}} \beta_{12}{\rm{sign}}
\beta_{22}\sqrt{p_{12}p_{22}} e^{j(\gamma_{12}-\gamma_{22})}\; ;$

or

$1 + \sigma K e^{j(\gamma_1-\gamma_2)}=0,$ 

where $K=\sqrt{p_{12}p_{22}}/$
$\sqrt{p_{11}p_{21}}$ and $\gamma_1=\gamma_{12}-\gamma_{22}, \gamma_2=\gamma_{11}-\gamma_{21}.$ 
Thus  

$\sinh(\gamma_1-\gamma_2)=0$ and
\begin{equation}
\label{PH}
\gamma_1=\gamma_2
\end{equation}
(we recall that in the standard quantum formalism we have $\gamma_1=\gamma_2+\pi\; (\mod 2\pi)).$
We also have 

$1+\sigma K \cosh(\gamma_1-\gamma_2)=0.$ 

Thus $\sigma=-1$ and $K=1.$ So sign-condition (\ref{SI}) is always satisfied for a
unitary matrix U=$(\beta_{ij}).$ 
The equality $K=1$ is equivalent to double stochasticity of the transition matrix of probabilities 
$P =(p_{ij}=|\beta_{ij}|^2).$ Therefore the matrix $U=(\beta_{ij})$ is a
{\bf{G}}-unitary matrix iff the corresponding matrix of probabilities
$P=(p_{ij})$ is a double stochastic matrix, $\sigma=-1,$ and hyperbolic phases satisfy to (\ref{PH}).

The H-quantum formalism (special calculus in a {\bf{G}}-linear space) represents probabilistic transformations

$q_1=p_1p_{11}+p_2p_{21}\pm 2\sqrt{p_1 p_2 p_{11} p_{21}} \cosh\theta\;,$

$q_2=p_1p_{12}+p_2p_{22}\mp 2\sqrt{p_1 p_2 p_{12} p_{22}} \cosh\theta\;,$ 

where $\theta=\gamma_{11}-\gamma_{21}=\gamma_{12}-\gamma_{22}.$

The situation is similar to the ordinary quantum formalism. However, there is 
the important difference between these formalisms. In the T-quantum formalism the condition of 
${{\bf{C}}}$-unitarity of $U=(\beta_{ij})$ was also sufficient to get
physically meaningful transformation of probabilities: all possible phases 
$\theta$ give meaningful probabilistic transformation for the fixed ${\bf{C}}$-unitary matrix 
$U=(\beta_{ij}).$ It is not so in the H-quantum formalism. 
The ${\bf{G}}$-unitary of $U=\beta_{ij}$ is not sufficient to get 
physically meaningful probabilities for all H-phases $\theta.$ 
Besides condition (\ref{NT}), we have also condition (\ref{E1}) 
which provides nonnegativity of probabilities $q_j=p_j^a=|\beta_j|^2.$ 

We set $t=p_{11}=p_{22}$ (so $p_{12}=p_{21}=1-t$), $0<t<1$ (we recall that $P=(p_{ij})$ 
is a double stochastic matrix). We also set $p_1=s, $ so $p_2=1-s, 0<s<1.$
Let us consider the case in that sign $\beta_{11}$ sign $\beta_{21}=-1.$ Hence 
sign $\beta_{12}$ sign $\beta_{22}=1.$ Here

$q_1=s t+(1-s)(1-t)-2\sqrt{s(1-s)t(1-t)} \cosh \theta\;,$ 

$q_2=s(1-t)+(1-s)t+2\sqrt{s(1-s)t(1-t)}\cosh \theta\;.$

\medskip

Thus $\cosh \theta\leq\frac{s t + (1-s)(1-t)}{2\sqrt{s(1-s)t(1-t)}}=e(s, t).$

Thus physical H-behaviour is possible only for probabilities $s, t$ such that $e(s, t)\geq 1$ 
(in the case of the equality H and T-behaviours coincide).

We note that there is no
an analogue of the superposition principle in the H-quantum formalism. 
{\bf{G}}-unitary transformations preserve normalization condition (\ref{NT}), 
but they do not preserve positivity conditions (\ref{E1}). 

We now turn back to the general case in that the $P$ need not be double stochastic.
We consider again equation (\ref{NT}) which is equivalent to (\ref{CON}) (with coefficients
belonging to hyperbolic algebra).
We have already studied the special class of solutions of equation (\ref{CON}) given by 
equation (\ref{CON1}). These solutions are given by ${{\bf{G}}}$-unitary matrixes. We now consider
the general equation:
\begin{equation}
\label{Eq3}
\sigma K \cosh(\eta + \gamma_2)+ \cosh (\eta + \gamma_1)=0.
\end{equation}
As $\sigma=-1, $ we finally get the equation 

$K \cosh \theta_2=-\cosh\theta_1$ 

(compare to (\ref{OT})).The presence of the H-phase 
$\eta=\xi_1-\xi_2$ plays the role of memory on the preparation procedure ${\cal E}$ 
(which produced an ensemble $S$ represented by the state $\varphi$).

We remark that equation (\ref{Eq3}) has following two solutions for $K=1$ (double stochastic matrix):

$\cosh(\eta + \gamma_2)=\cosh(\eta + \gamma_1) \rightarrow \eta + \gamma_2=\eta + 
\gamma_1$ or $\eta + \gamma_2=-\eta -\gamma_1.$ 

In the first case we have the H-quantum 
solution,$ \gamma_1= \gamma_2,$
and in the second case we have a new solution, $2\eta+\gamma_2+\gamma_1=0,$
that corresponds to non unitary transition matrix $U.$

\section{Conclusions}

Our frequency analysis of probabilities related to transitions from one
experimental arrangement (context, complex of physical conditions) to another showed:

1. "Quantum rule" for interference of probabilistic alternatives can be obtained
in purely contextualists approach; in particular, without to apply to wave arguments.

2. Both "quantum" and "classical" probabilities can be interpreted as frequency 
probabilities. Specific "quantum behaviour" of probabilities in experiments with 
quantum particles is related to the specific relation between elementary particles
and experimental arrangement. There in "quantum world" transition from one context
to another produces statistical perturbations that change classical Bayes'
formula (by the  additive interference term). In "classical world" such perturbations
are negligibly small statisticaly.

3. Transformations of probabilities corresponding to context transitions can be classified
according to the magnitudes of statistical perturbations: trigonometric, hyperbolic,
hyper-trigonometric. In particular, contextual modifications of classical Bayes' formula 
are not reduced to "quantum rule" for interference of probabilistic alternatives. 
There exists non classical/quantum trigonometric interference of probabilistic alternatives
as well as hyperbolic interference.

4. The main distinguishing feature of "quantum probabilistic transformations"
is not the appearance of the $\cos \theta$-interference term, but the double
stochasticity of the matrix of transition probabilities.

5. Starting with the trigonometric transformation of probabilities, we 
get (with the aid of $\cos$-theorem) complex amplitude representation 
of contextual probabilities. This gives the possibility to construct
a complex linear space representation of contextual probabilistic
calculus.  In general, we could not represent a trigonometric probabilistic
transformation in a complex Hilbert space. It is possible only for
double stochastic matrixes of transition  probabilities (that corresponds
to unitary transformations of a Hilbert space).

6. One of special features of general ${\bf C}$-linear representation
of contextual probabilities is the violation of the superposition principle.
It seems that this fundamental principle is a consequence 
of double stochasticity.

7. Hyperbolic probabilistic transformations can be represented as linear
transformations in modules over the system (commutative algebra) ${\bf G}$
of hyperbolic numbers. 

8. One of special features of general ${\bf G}$-linear representation
of contextual probabilities is the violation of the superposition principle. In 
this case even double stochasticity of the matrix of transition probability
(${\bf G}$-unitarity of the corresponding ${\bf G}$-linear transformation)
does not imply the superposition principle.

9. Trigonometric transformations correspond to  context transitions inducing
relatively small statistical perturbations; hyperbolic - relatively large.

10. In principle, non classical/quantum probabilistic behaviour (trigonometric
as well as hyperbolic) could be simulated numerically, see [32].

Finally, we make a remark on the contextualists viewpoint to superselection rules.
Superselection rules are closely related to the superposition principle. With a superselection 
rule unitarity (double stochasticity of the matrix of transition probabilities)
or linear combinations do not imply coherent superposition. Superselection rules also are important
since they are relevant to macroscopic quantum systems [33], [34]. 

I think that superselection rules give restrictions on physical realization of
some preparation procedures, namely filtration (selection) procedures
that give a possibility to transform an ensemble of physical systems
prepared under one fixed complex of conditions ${\cal S}$ 
into an ensemble of physical systems prepared under some special complex of physical
conditions ${\cal S}^\prime.$ There exists  complexes ${\cal S}$ and ${\cal S}^\prime$ such that
it is impossible to create the corresponding transformation procedure. 
However, I think (and it may be that I am wrong) that superselection
rules could not be analysed in general probabilistic framework. Each rule is closely
connected to some fixed class of physical systems under consideration.
If we represent in the same linear (in particular, Hilbert) space contextual
probabilities for distinct classes of physical systems, then we shall get distinct
classes of contexts that could not be transformed into each other.

I would like to thank L. Ballentine, S. Albeverio, E. Beltrametti,
T. Hida, D. Greenberger, S. Gudder, I. Volovich, W. De Muynck, J. Summhammer, P. Lahti, J-A. Larsson, H. Atmanspacher, 
B. Coecke, S. Aerts, A. Peres, A. Holevo,  E. Loubenets,  L. Polley, A. Zeilinger, C. Fuchs, R. Gill, L. Hardy,
B. Hiley, S. Goldshtein, A. Plotnitsky, A. Shimony, R. Jozsa, J. Bub, C. Caves, K. Gustafsson, H. Bernstein
for fruitful (and rather critical) discussions.

{\bf References}

[1] P. A. M.  Dirac, {\it The Principles of Quantum Mechanics}
(Claredon Press, Oxford, 1995).

[2] J. von Neumann, {\it Mathematical foundations
of quantum mechanics} (Princeton Univ. Press, Princeton, N.J., 1955).

[3] W. Heisenberg, {\it Physical principles of quantum theory.}
(Chicago Univ. Press, 1930).

[4] R. Feynman and A. Hibbs, {\it Quantum Mechanics and Path Integrals}
(McGraw-Hill, New-York, 1965).

[5] J. M. Jauch, {\it Foundations of Quantum Mechanics} (Addison-Wesley, Reading, Mass., 1968).

[6]  P. Busch, M. Grabowski, P. Lahti, {\it Operational Quantum Physics}
(Springer Verlag, 1995).

[7] B. d'Espagnat, {\em Veiled Reality. An anlysis of present-day
quantum mechanical concepts} (Addison-Wesley, 1995). 

[8] A. Peres, {\em Quantum Theory: Concepts and Methods} (Kluwer Academic Publishers, 1994).

[9] E. Beltrametti  and G. Cassinelli, {\it The logic of Quantum mechanics.}
(Addison-Wesley, Reading, Mass., 1981).

[10]  L. E. Ballentine, {\it Quantum mechanics} (Englewood Cliffs, 
New Jersey, 1989).

[11] N. Bohr, {\it Phys. Rev.,} {\bf 48}, 696-702 (1935).

[12] L. Accardi, The probabilistic roots of the quantum mechanical paradoxes.
{\em The wave--particle dualism.  A tribute to Louis de Broglie on his 90th 
Birthday,} ed. S. Diner, D. Fargue, G. Lochak and F. Selleri
(D. Reidel Publ. Company, Dordrecht, 297--330, 1984);

L. Accardi, {\it Urne e Camaleoni: Dialogo sulla realta,
le leggi del caso e la teoria quantistica.} Il Saggiatore, Rome (1997).

[13] L. E. Ballentine, {\it Rev. Mod. Phys.}, {\bf 42}, 358--381 (1970).

[14]  S.P. Gudder,  J. Math Phys., {\bf 25}, 2397- 2401 (1984).

[15]  W. De Muynck, W. De Baere, H. Marten,
 Found. of Physics, {\bf 24}, 1589--1663 (1994);

W. De Muynck, J.T. Stekelenborg,  Annalen der Physik, {\bf 45},
N.7, 222-234 (1988).

[16] I. Pitowsky,  Phys. Rev. Lett, {\bf 48}, N.10, 1299-1302 (1982);
 Phys. Rev. D, {\bf 27}, N.10, 2316-2326 (1983).

[17] E. B. Davies, {\it Quantum theory of open systems} (Academic Press, 
London, 1976).

[18] G. Ludwig, {\it Foundations of quantum mechanics} (Springer, 
Berlin, 1983).

[19] A.Yu. Khrennikov, {\it Interpretations of 
probability} (VSP Int. Publ., Utrecht, 1999).

[20]  A. Yu. Khrennikov,  {\it J. of Math. Physics}, {\bf 41}, N.9, 5934-5944 (2000).

[21] A. Yu. Khrennikov,  {\it Phys. Lett.},
A, 278, 307-314 (2001).

[22] A. Yu. Khrennikov, {\it Ensemble fluctuations and the origin of quantum probabilistic
rule.} Rep. MSI, V\"axj\"o Univ., {\bf 90}, October (2000).

[23] A. Yu. Khrennikov, {\it Hyperbolic Quantum Mechanics.} Preprint: quant-ph/0101002, 31 Dec 2000.
 
[24] A. S. Holevo, {\it Probabilistic and statistical aspects of quantum 
theory.} North-Holland, Amsterdam (1982).

[25] R.  von Mises, {\it The mathematical theory of probability and
 statistics}. Academic, London (1964); Ville J., {\it Etude critique de la notion de collective}, Gauthier--
Villars, Paris (1939); E. Tornier , {\it Wahrscheinlichkeitsrechnunug und allgemeine 
Integrationstheorie.}, Univ. Press, Leipzing (1936).

[26] A. N. Kolmogoroff, {\it Grundbegriffe der Wahrscheinlichkeitsrech}
Springer Verlag, Berlin (1933); reprinted:
{\it Foundations of the Probability Theory}. 
Chelsea Publ. Comp., New York (1956);

[27] L. Ballentine, Interpretations of probability and quantum theory. {\it Quantum Probability
and Related Fields,} To be published.

[28] E. Wigner, {\it Quantum -mechanical distribution functions revisted,}
in: {\it Perspectives in quantum theory}.  Yourgrau W. and van der Merwe A., editors,
MIT Press, Cambridge MA (1971).

[29]  P. A. M. Dirac,{\it Proc. Roy. Soc. London}, {\bf A 180}, 1--39 (1942).

[30]   R. P. Feynman, Negative probability. {\it Quantum Implications,  
Essays in Honour of David Bohm}, B.J. Hiley and F.D. Peat, editors,
Routledge and Kegan Paul, London, 235--246 (1987).

[31]  A. Yu. Khrennikov, {\it Supernalysis.}  (Kluwer Academic Publishers, 
Dordreht/Boston/London, 1999).

[32] A. Khrennikov, {\it Quantum statistics via perturbation effects of preparation procedures.}
Preprint: quant-ph/0103065, 13 March 2001.

[33] K.K. Wan, F. Harrison, {\it Phys. Lett.} A, {\bf 174}, 1 (1993)

[34]  K.K. Wan, J. Bradshaw, C. Trueman,  F. Harrison,  {\it Found. Phys.},
{\bf 28},1739-1783 (1998).

\end{document}